\documentclass[final,3p,times,twocolumn]{elsarticle}

\usepackage{graphicx}

\usepackage{amssymb}

\usepackage{hhline}
\usepackage{epsfig}
\usepackage{amsmath}
\usepackage{dcolumn}
\usepackage{multirow}
\usepackage{array}
\usepackage{color}

\journal{Physics Letters B}

\newcommand{\be}{\begin{equation}}
\newcommand{\ee}[1]{\label{#1} \end{equation}}
\newcommand{\ba}{\begin{eqnarray}}
\newcommand{\ea}[1]{\label{#1} \end{eqnarray}}
\newcommand{\nl}{\nonumber \\}

\begin{document}

\begin{frontmatter}


 \ead{karoly.uermoessy@cern.ch}

\title{Disentangling Soft and Hard Hadron Yields\\in $PbPb$ Collisions at $\sqrt{s_{NN}}$ = 2.76 ATeV}

\author{K. Urmossy$^{\,a}$, T.~S.~Bir\'o$^{\,a}$, G. G. Barnaf\"oldi$^{\,a}$ and Z. Xu$^{\,b}$}

\address[label1]{Wigner Research Center for Physics of the Hungarian Academy of Sciences\nl29--33 Konkoly--Thege Mikl\'os Str., H-1121 Budapest, Hungary\vspace{2mm}}

\address[label2]{Physics Department, Brookhaven National Laboratory\\
Upton, NY 11973, USA}


\begin{abstract}

We demonstrate that charged pion spectra in central and peripheral $PbPb$ collisions at $\sqrt s$ = 2.76 ATeV obtained via perturbative quantum chromodynamics improved parton model calculations \cite{bib:pQCDGusty} can be approximated by the Tsallis distribution for transverse momenta $p_T \geq$ 4 GeV/c. Then, we propose a model in which, hadrons produced in heavy-ion collisions stem either from ``soft'' or ``hard'' processes, and are distributed according to the Tsallis distribution in both types of yields. We describe transverse spectra \textit{as well as} azimuthal anisotropy ($v_2$) of charged hadrons stemming from various centrality $PbPb$ collisions at $\sqrt s$ = 2.76 ATeV \cite{bib:cmsdNdpT,bib:cmsv2,bib:ALICEdNdpT} \textit{analytically}. 
In addition, we find that the anisotropy decreases for more central collisions.

\end{abstract}


\end{frontmatter}



\date{\today}

\section{Introduction}
\label{sec:intro}

\begin{figure}
\begin{center}
\includegraphics[width=0.45\textwidth, height=0.34\textheight]{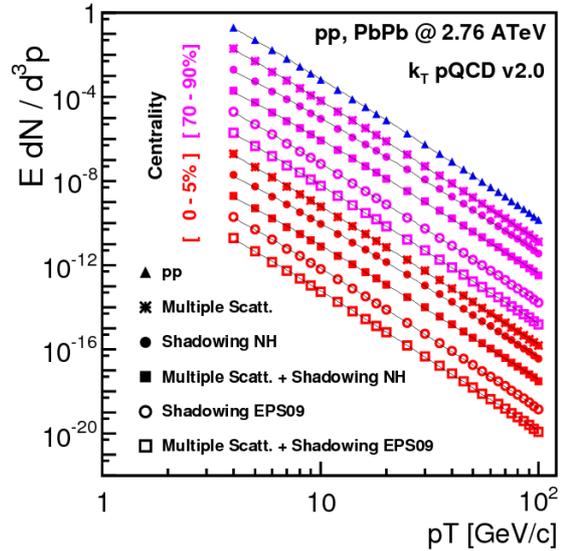}
\end{center}
\caption{$pQCD$-improved parton model results for charged pion spectra in $pp$ and central ([0--5\%]) as well as peripheral ([70--90\%]) $PbPb$ collisions at $\sqrt s$ = 2.76 ATeV collision energy (shifted by factors of 0.1). In these calculations, various combinations were used, in which multipliple scattering and shadowing (\textit{``NH''} and \textit{``EPS09''}) were turned on/off. The theoretical framework is discussed in \cite{bib:pQCDGusty}. \textbf{Solid curves} are fitted Tsallis distributions. \label{fig:pQCD}}
\end{figure}

Because of its short lifetime, the only way to examine the quark-gluon plasma (QGP) formed in ultra-relativistic heavy-ion collisions (HIC), is looking at the particles stemming from it. Spectra, angular correlations and their dependence on the circumstances of the collision can then be studied. These distributions are effected by hadron yields stemming not only from the QGP (we refer to as ``soft'' yields), but also from jets (we call ``hard'' yields). Though, due to jet-medium interactions, it might be a challenging task to separate particles originating from jets and the QGP, for a first approximation, we make out hadron spectra in HICs as
\be
p^0\frac{dN}{d^3\mathbf{p}} =  p^0\frac{dN}{d^3\mathbf{p}}^{\textrm{hard}} + p^0\frac{dN}{d^3\mathbf{p}}^{\textrm{soft}} \;. 
\ee{eq1}
That is, yields stemming from semi-hard or mixed processes or mini-jets are assumed to be included in either the soft or the hard contribution.

For the description of hard hadron yields, we will use the Tsallis distribution, because it provides a reasonably good approximation for the transverse spectra of charged pions stemming from central as well as peripheral $PbPb$ collisions obtained via perturbative quantum chromodynamics ($pQCD$) improved parton model calculations (cf Fig.~\ref{fig:pQCD}). As the Tsallis distribution also describes measured transverse spectra of charged and identified hadrons \cite{bib:phenixPP}--\cite{bib:Wong2} as well as the spectrum obtained by parton model calculations \cite{bib:BGG_FF,bib:pQCD}, in the case of proton-proton collisions (as argued in \cite{bib:Wong1,bib:Wong2}), we conclude that the effects of shadowing and multiple scattering in heavy-ion collisions, are reflected in the modification of the Tsallis distribution's parameters. The modified parameters can be determined by fitting the high-transverse momentum ($p_T\geq$ 6 GeV/c) part of the measured spectra in HICs. 

After the determination and subtraction of the hard part, we expect to be left with the soft part of the spectrum. This, we identify as yields stemming dominantly from the QGP. The soft spectrum is also found to fit a Tsallis distribution, but with parameters different from those of other models in the literature \cite{bib:Wilk2}--\cite{bib:Wilk7}. In those models, the hard part of the spectrum has not been subtracted. For the emergence of the Tsallis distribution in the soft part of the spectrum, there is a chance to bring statistical arguments based on non-extensive thermodynamics \cite{bib:BiroEPJA40,bib:BiroJako}, or on super-statistics \cite{bib:UKppNdep,bib:Wilk3,bib:Wilk4,bib:Wilk5,bib:Wilk6,bib:Wilk7,bib:UKppFF,bib:UKeeFF}.

This ``soft'' + ``hard'' type model using the Tsallis distribution \textit{simultanously} describes the transverse spectra \textit{and} the azimuthal anisotropy ($v_2$) of charged hadrons stemming from $PbPb$ collisions at $\sqrt s$ = 2.76 ATeV. The parameters obtained from fits to the spectra and to $v_2$ are consistent with each other. As for the centrality dependence of the fitted parameters, CMS \cite{bib:cmsdNdpT,bib:cmsv2} and ALICE \cite{bib:ALICEdNdpT} measurements suggest only slightly different behaviour.



We note that transverse spectra and $v_2$ of various identified hadrons measured at RHIC energy have been described by a similar model \cite{bib:Tang1,bib:Tang2,bib:Shao}. In that model, spectra measured in $pp$ collisions have been used as hard yields, and it has been conjectured that hard yields are suppressed at low $p_T$.

In Sec.~\ref{sec:vn}, calculation of transverse hadron spectra and $v_n$ are shown. Secs.~\ref{sec:fitspec}~and~\ref{sec:fitv2} contain fits to charged hadron spectra and $v_2$ measured in various centrality $PbPb$ collisions at $\sqrt s$ = 2.76 ATeV by the CMS \cite{bib:cmsdNdpT,bib:cmsv2} and the ALICE \cite{bib:ALICEdNdpT} collaborations. Summary is given in Sec.~\ref{sec:conc}.

\section{Transverse Spectrum and $v_n$}
\label{sec:vn}

As possibly the simplest approximation, we obtain the transverse spectrum as a sum of hadrons with momentum $p^{\,\mu}$, coming from sources flying with velocities $u^{\,\mu}$ as
\be
\left.p^0\,\frac{dN}{d^3p}\right|_{y=0} = \int \limits_{-\infty}^{+\infty} d\zeta \int \limits_0^{2\pi} d\alpha \, f\big[ u_\mu p^\mu \big] \;.
\ee{eq2}
Here, $\alpha$ is the azimuth angle and $\zeta = \frac{1}{2}\ln[(t+z)/(t-z)]$. We parametrize hadron momenta as
\be
p^{\,\mu} = (m_T\cosh y,m_T\sinh y,p_T \cos\varphi, p_T\sin\varphi)\;,
\ee{eq3}
with $y = \frac{1}{2}\ln[(p^0+p^z)/(p^0-p^z)]$ and $\varphi$ being the azimuth angle of the hadron momentum. We choose a cylindrically symmetric radial flow
\be
u^{\,\mu} = (\gamma\cosh\zeta,\gamma\sinh\zeta,\gamma v\cos\alpha, \gamma v \sin\alpha)\;,
\ee{eq4}
with $\gamma = 1/\sqrt{1-v^2}$, and assume that $v$ depends only on $\alpha$. Though, it is assumed that in each source, the momentum  distribution of hadrons $f$ is a function of the co-moving energy
\be
\left. u_\mu p^\mu \right|_{y=0} = \gamma \big[m_T\cosh\zeta - v p_T \cos(\varphi-\alpha) \big]\;,
\ee{eq5}
the sources may be fireballs \cite{bib:OldBoys1}--\cite{bib:OldBoys3}, clusters \cite{bib:Wibig2,bib:Wibig,bib:Wibig3,bib:Becattini11,bib:Liu1} or even jets \cite{bib:UKppFF,bib:UKeeFF}.

We write the transverse flow as a series,
\be
v(\alpha)= v_0 + \sum\limits_{m=1}^\infty \delta v_m \cos(m\alpha) \equiv v_0 + \delta v(\alpha) \;,
\ee{eq6}
and suppose that $\delta v(\alpha) << 1$. We use the Taylor expansion 
\be
\left.f\big[u_\mu p^\mu \big]\right|_{y=0} = \sum_{m=0}^\infty \frac{ \left[\delta v(\alpha)\right]^m}{m!} \frac{\partial^m}{\partial v_0^m}\, \left.f\big[ u_\mu p^\mu \big]\right|_{\,y\,=\,0}^{\,v(\alpha)\,=\,v_0}\;,
\ee{eq7}
and keep only the leading non-vanishing terms in $\delta v(\alpha)$.

Provided that $f$ is a rapidly decreasing function, we approximate integrals with respect to $\zeta$ and $\varphi$ by the maximal value of the integrands times the integration interval. Thus, the $\varphi$ integrated transverse spectrum becomes
\ba
\left. \frac{dN}{2\pi p_T dp_T dy} \right|_{\,y\,=\,0} &=& \int \limits_0^{2\pi} \frac{d\varphi}{2\pi}  \left.p^0\,\dfrac{dN}{d^3p}\right|_{\,y\,=\,0} \;=\; \nl
&=& \sum_{m=0}^\infty \frac{a_m}{m!} \frac{\partial^m}{\partial v_0^m}\, f\big[E(v_0)\big] \;\approx\; \nl
&\approx& f\big[E(v_0)\big] \;+\; \mathcal{O}\left(\delta v^2\right)\;,\nl
\ea{eq8}
with $E(v_0) = \gamma_0 (m_T - v_0 p_T)$ and $a_m = \int\limits_0^{2\pi} d\alpha \,\left[\delta v(\alpha)\right]^m$. Similarly, the azimuthal anisotropy becomes
\ba
v_n &=& \frac{\int \limits_0^{2\pi} d\varphi \cos(n\varphi) \, \left.p^0\,\dfrac{dN}{d^3p}\right|_{y=0}}{ \int \limits_0^{2\pi} d\varphi \, \left.p^0\,\dfrac{dN}{d^3p}\right|_{y=0} } \;\approx\;\nl
&\approx& \frac{\delta v_n \gamma^3_0}{2} \frac{(v_0\, m_T - p_T) f'\big[E(v_0)\big]}{f\big[E(v_0)\big]} + \mathcal{O}\left(\delta v^2\right)\;,\nl
\ea{eq9}
with $\delta v_n$ defined in Eq.~(\ref{eq6}).

For example, in the case of the Boltzmann-distribution $f \sim \exp\big[-\beta E(v_0)\big]$, the anisotropy is
\be
v^{\textrm{BG}}_n \approx \frac{\delta v_n \,\beta\, \gamma^3_0}{2} (p_T - v_0\, m_T) \; +\; \mathcal{O}\left(\delta v^2\right)\;.
\ee{eq10}
Thus, $v^{\textrm{BG}}_n \propto p_T$ if $p_T>>m$.

In the case of the Tsallis distribution, $f \sim [1 + (q-1)\,\beta\, E(v_0)]^{-1/(q-1)}$, the anisotropy
\be
v^{\textrm{TS}}_n \approx \frac{\delta v_n \,\beta\, \gamma^3_0}{2} \frac{p_T - v_0\, m_T}{1 + (q-1)\,\beta\,\gamma_0(m_T - v_0\,p_T)} \; +\; \mathcal{O}\left(\delta v^2\right) \;.
\ee{eq11}
Thus, $v^{\textrm{TS}}_n$ saturates when $(q-1)\,\beta\,\gamma_0 (1-v_0) p_T >> 1$.

\section{Fits to the Spectrum of Charged Hadrons in $PbPb$ Collisions at $\sqrt{s_{NN}}$ = 2.76 ATeV}
\label{sec:fitspec}

As conjectured in Sec.~\ref{sec:intro}, we make out the transverse spectrum of charged hadrons by the sum of hard and soft yields
\be
\left. \frac{dN}{2\pi p_T dp_T dy} \right|_{y=0} = f_{hard} + f_{soft}\;,
\ee{eq12}
where both contributions are assumed to be Tsallis-distributions
\be
f_i = A_i \left[1 + \frac{(q_i-1)}{T_i}[\gamma_i(m_T-v_i p_T) - m] \right]^{-1/(q_i-1)}
\ee{eq13}
($i$ = soft or hard). These yields have maxima at $p^{max}_{T} = \gamma_i\, m\, v_i$. As long as these maxima are below the measurement range, which is the case in this analysis, the isotropic part of the transverse flow, $v_i$ (denoted by $v_0$ in Eq.~(\ref{eq6}) in Sec.~\ref{sec:vn}) cannot be determined accurately. As the dominant part of charged hadrons consists of pions, the argument of $f_i$-s may be approximated by $[\gamma_i(m_T - v_i p_T) -m]/T_i \approx p_T/T^{Dop}_i$ with the Doppler-shifted parameters
\be
T^{Dop}_i = T_i\,\sqrt{\frac{1+v_i}{1-v_i}} \;.
\ee{eq14}

\begin{figure}
\begin{center}
\includegraphics[width=0.45\textwidth, height=0.34\textheight]{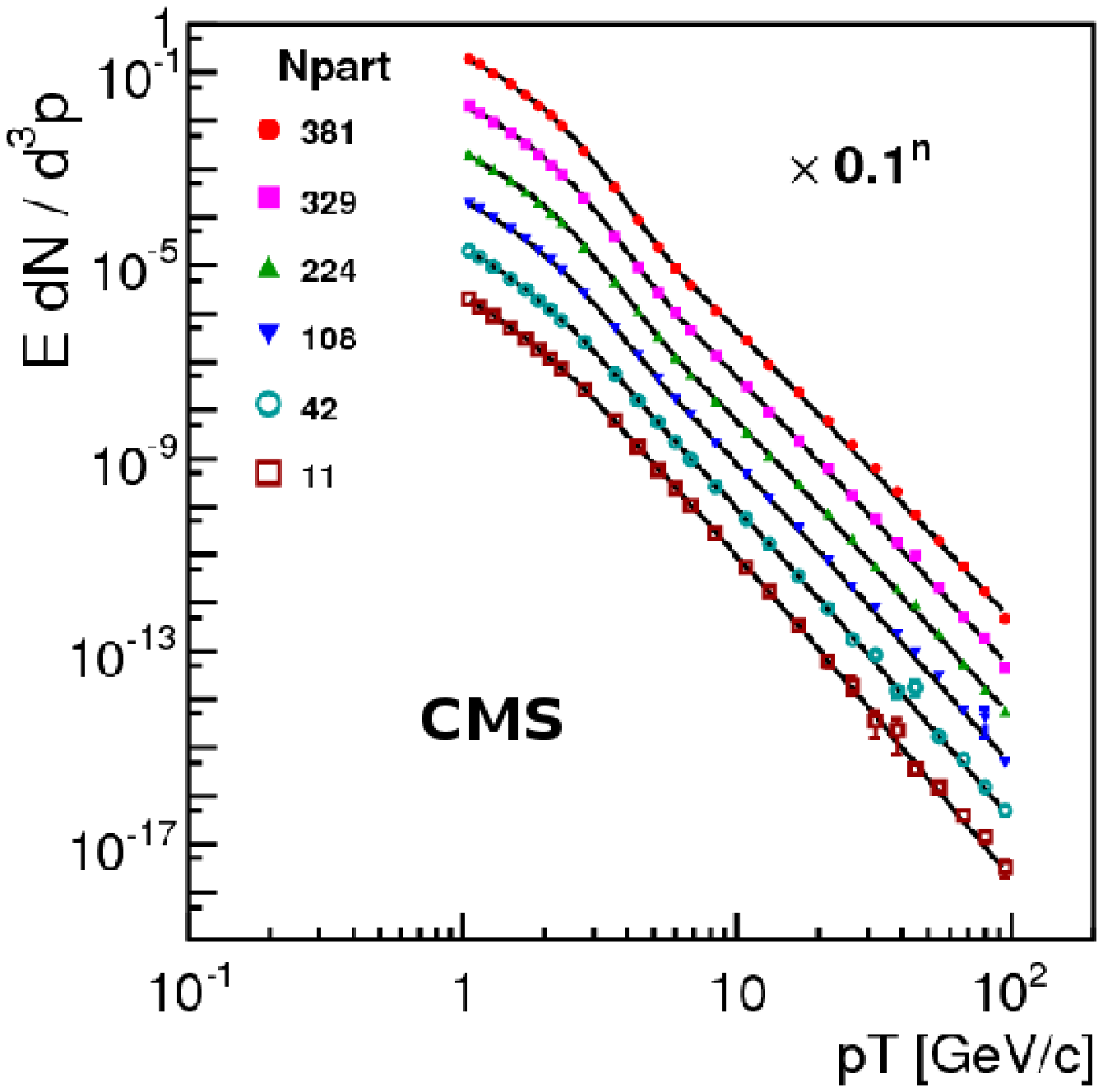}
\includegraphics[width=0.45\textwidth]{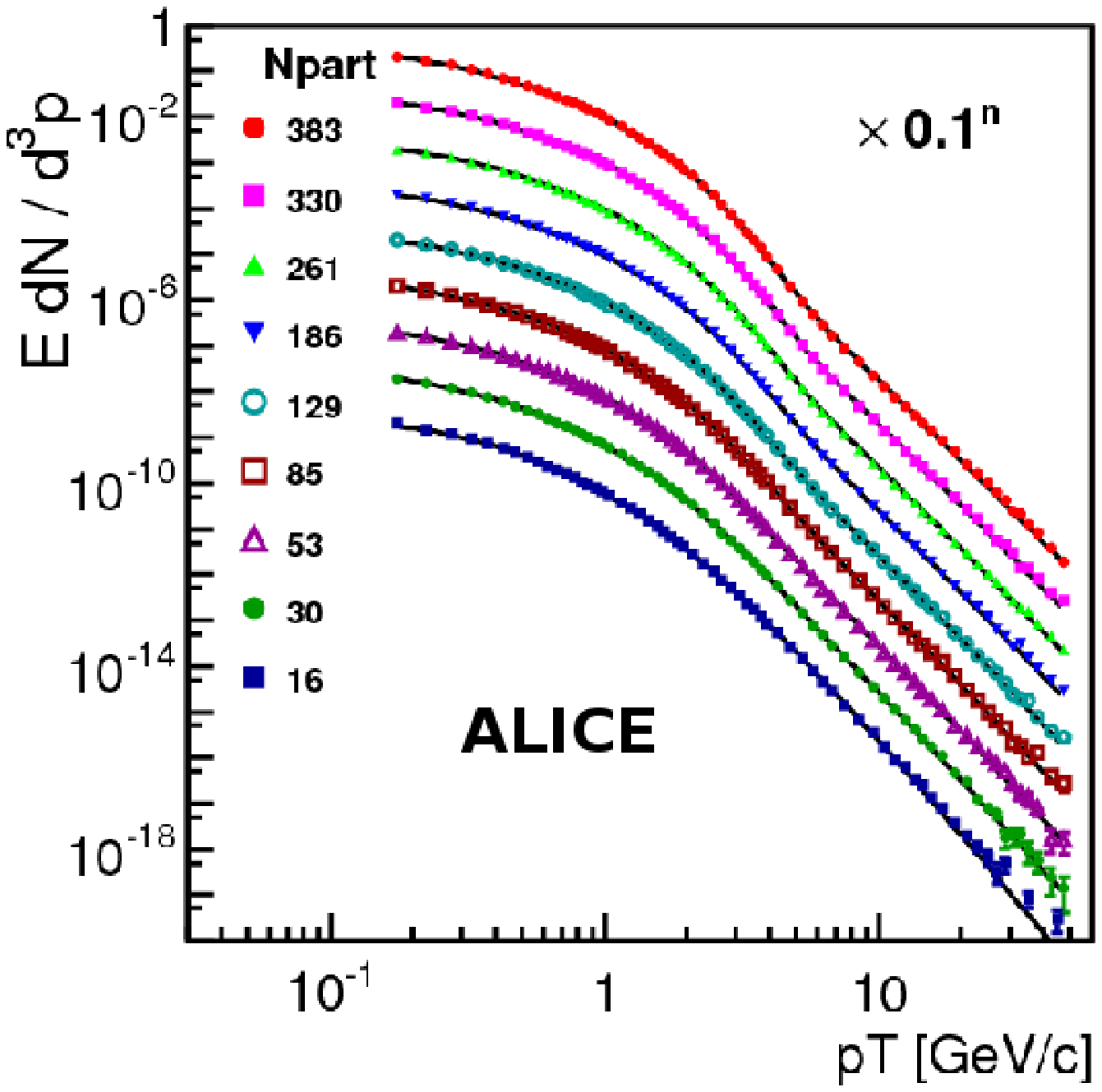}
\includegraphics[width=0.45\textwidth]{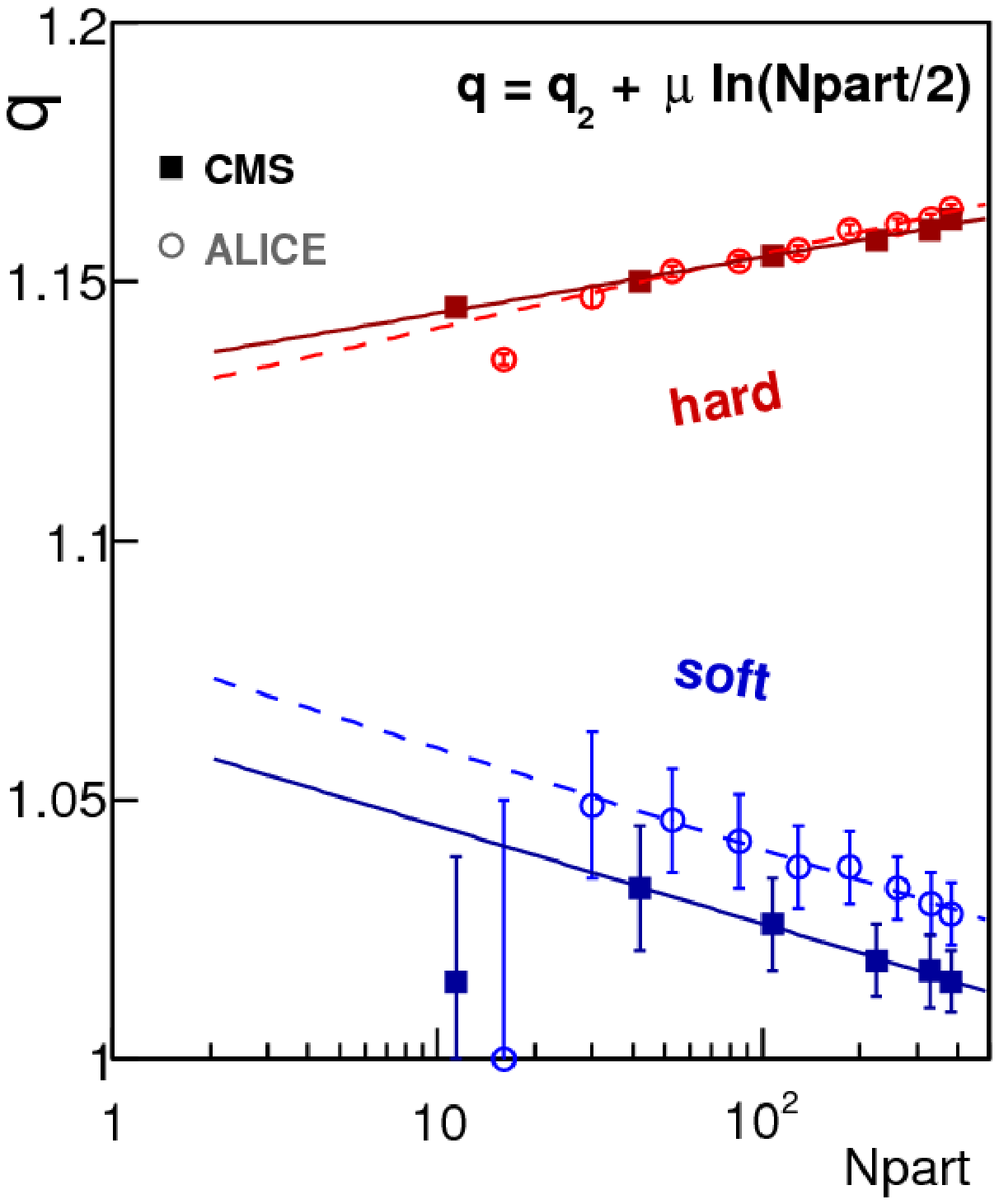} 
\includegraphics[width=0.45\textwidth]{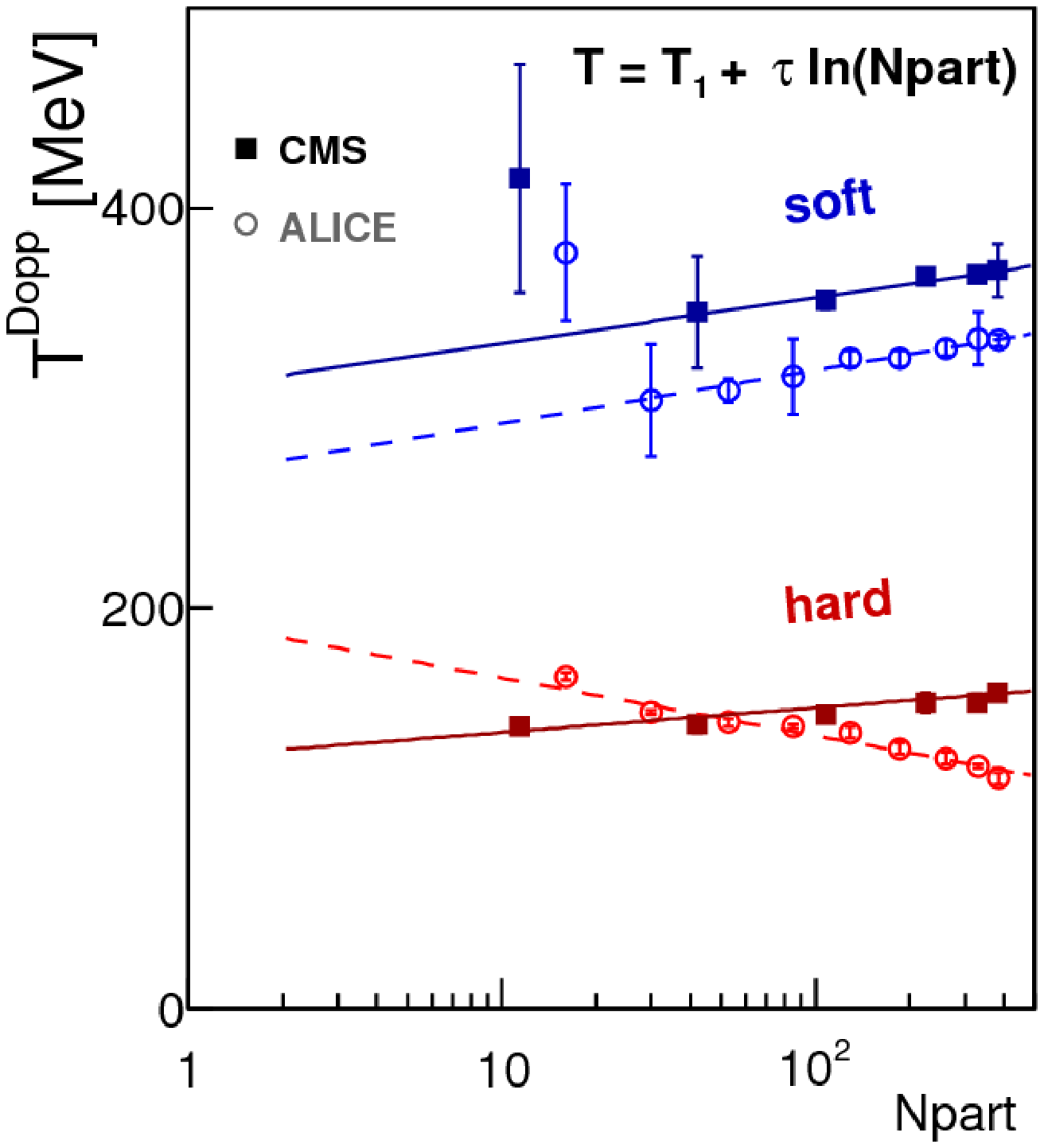} 
\end{center}
\caption{\textbf{Top,} transverse spectra of charged hadrons stemming from various centrality $PbPb$ collisions at $\sqrt s$ = 2.76 ATeV measured by the CMS \cite{bib:cmsdNdpT} (\textbf{left}) and ALICE \cite{bib:ALICEdNdpT} (\textbf{right}) Collaborations. Curves are fits of Eqs.~(\ref{eq12})~--~(\ref{eq14}). \textbf{Bottom,} centrality dependence of the fitted $q$ (\textbf{left}) and $T^{Dop}$ (\textbf{right}) parameters of soft and hard yields in Eqs.~(\ref{eq12})~--~(\ref{eq14}). Fitted parameters are enlisted in Tables~\ref{tab:CMSparam}--\ref{tab:ALICEparam}. Straight lines are in Eq.~(\ref{eq15}) with parameters enlisted in Table~\ref{tab:qTfit}. \label{fig:dNdpT}}
\end{figure}

As can be seen in the top panels of Fig.~\ref{fig:dNdpT}, Eqs.~(\ref{eq12})~--~(\ref{eq14}) describe CMS \cite{bib:cmsdNdpT} and ALICE \cite{bib:ALICEdNdpT} data on transverse spectra of charged hadrons stemming from $PbPb$ collisions of various centralities. Fitted parameters are shown in Tables~\ref{tab:CMSparam}~--~\ref{tab:ALICEparam} and in the bottom panels of Fig.~\ref{fig:dNdpT}. The dependence of the $q$ and $T^{Dop}$ parameters of the soft and hard yields on the event centrality (number of participating nucleons $N_{part}$) can be fitted by
\ba
q_i &=& q_{2,\,i} + \mu_i \ln(N_{part}/2) \;,\nl
T^{Dop}_i &=& T_{1,\,i} + \tau_i \ln(N_{part}) \;.
\ea{eq15}
Though the actual value of the transverse flow velocity cannot be determined in this model from the spectra of charged hadrons, it may be guessed using the value of the QGP-hadronic matter transition temperature obtained from lattice-QCD calculations. As the values of fitted $T^{Dop}_{soft}$ scatter around 340 MeV, in case of a flow velocity of $v_{soft} \approx$  0.6, the real $T_{soft}$ values would scatter around 170 MeV, which is close to the lattice result obtained e.g. in \cite{bib:lQCD}.



While the tendencies of how fit parameters depend on $N_{part}$ are similar, they are not the same within errors in the case of CMS \cite{bib:cmsdNdpT} and ALICE \cite{bib:ALICEdNdpT} measurements. It is to be noted that in \cite{bib:ALICEdNdpT}, centrality is determined using the distribution of hits in the VZERO detector, which has a rapidity coverage of 2.8 $\leq\eta\leq$ 5.1 and -3.7 $\leq\eta\leq$ -1.7. In the meanwhile, in \cite{bib:cmsdNdpT,bib:cmsv2}, the collision event centrality is determined from the event-by-event total energy deposition in both Hadron Forward calorimeters having rapidity coverage of 2.9 $\leq|\eta|\leq$ 5.2.

\section{Fits to $v_2$ of Charged Hadrons in $PbPb$ Collisions at $\sqrt{s_{NN}}$ = 2.76 ATeV}
\label{sec:fitv2}

As seen from Sec.~\ref{sec:vn}, up to $\mathcal{O}\left(\delta v^2 \right)$, the transverse spectrum in Eqs.~(\ref{eq12})~--~(\ref{eq13}) results in an azimuthal anisotropy of
\be
v_2 = \frac{w_{hard} \,f_{hard} + w_{soft}\, f_{soft}}{f_{hard} + f_{soft}}\;, 
\ee{eq17}
where the coefficient functions are
\be
w_i = \frac{\delta v_i\, \gamma^3_i}{2T_i} \frac{p_T - v_i\, m_T}{1 + \dfrac{q_i-1}{T_i}\,\big[\gamma_i(m_T - v_i\,p_T) - m \big]} \;.
\ee{eq18}
Again, $i=$ soft or hard, $v_i$ are the isotropic part of the transverse flow (denoted by $v_0$ in Eq.~(\ref{eq6}) in Sec.~\ref{sec:vn}). And $\delta v_i$ are the coefficients of $\cos(2\alpha)$ (denoted by $\delta v_2$ in Eq.~(\ref{eq6}) in Sec.~\ref{sec:vn}).

\begin{figure}
\begin{center}
\includegraphics[width=0.45\textwidth]{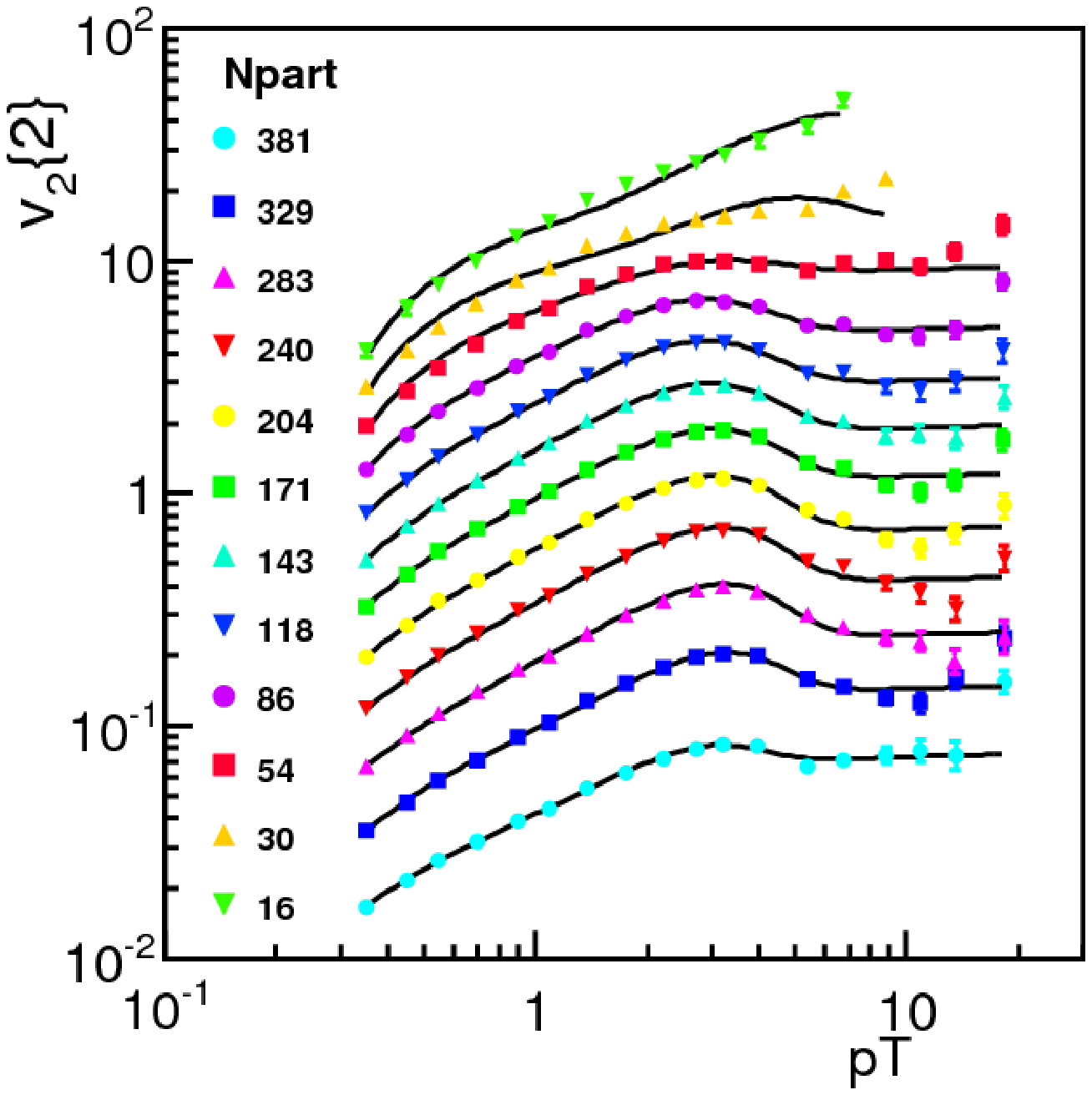} 
\includegraphics[width=0.45\textwidth]{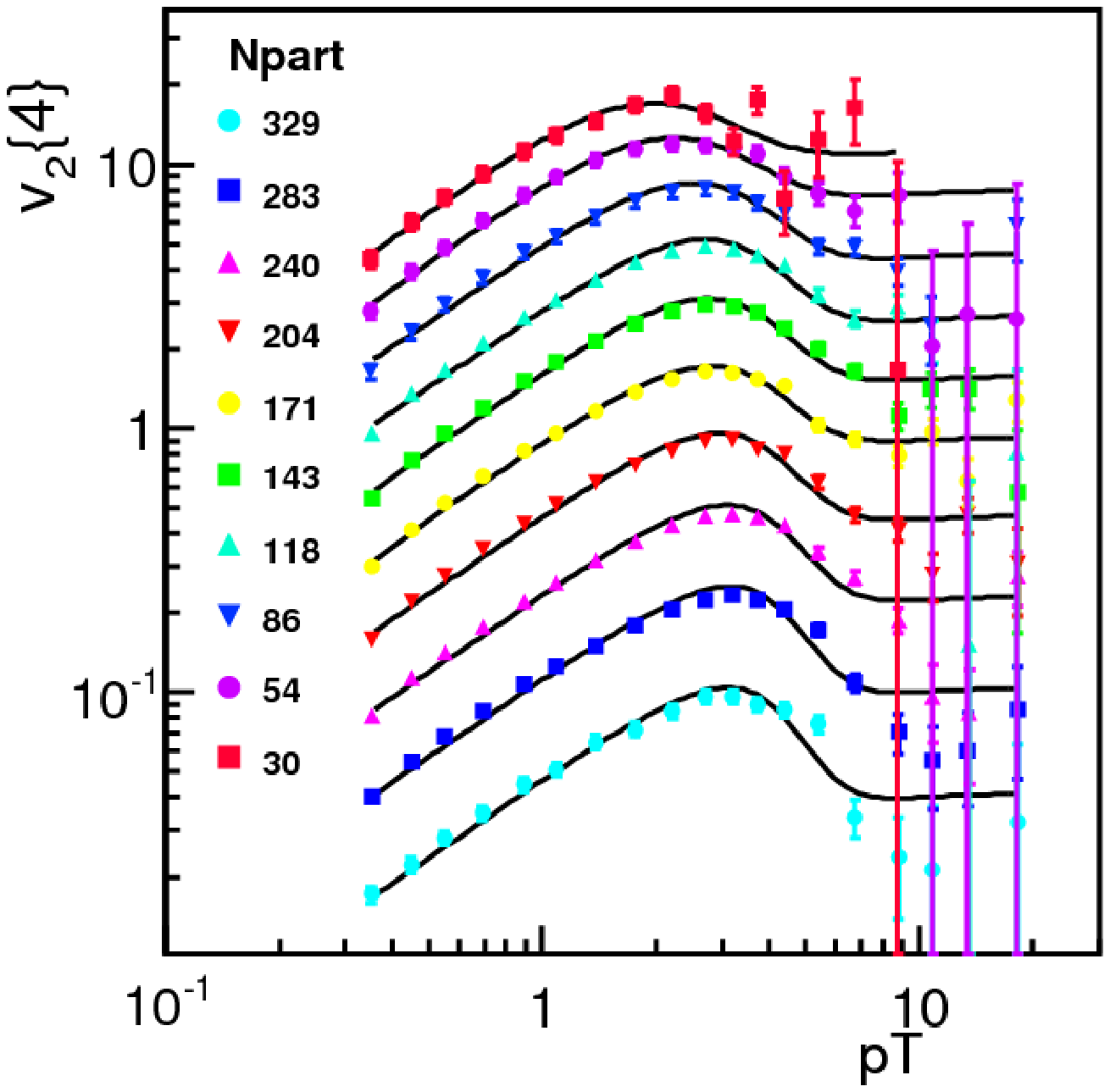} 
\includegraphics[width=0.45\textwidth]{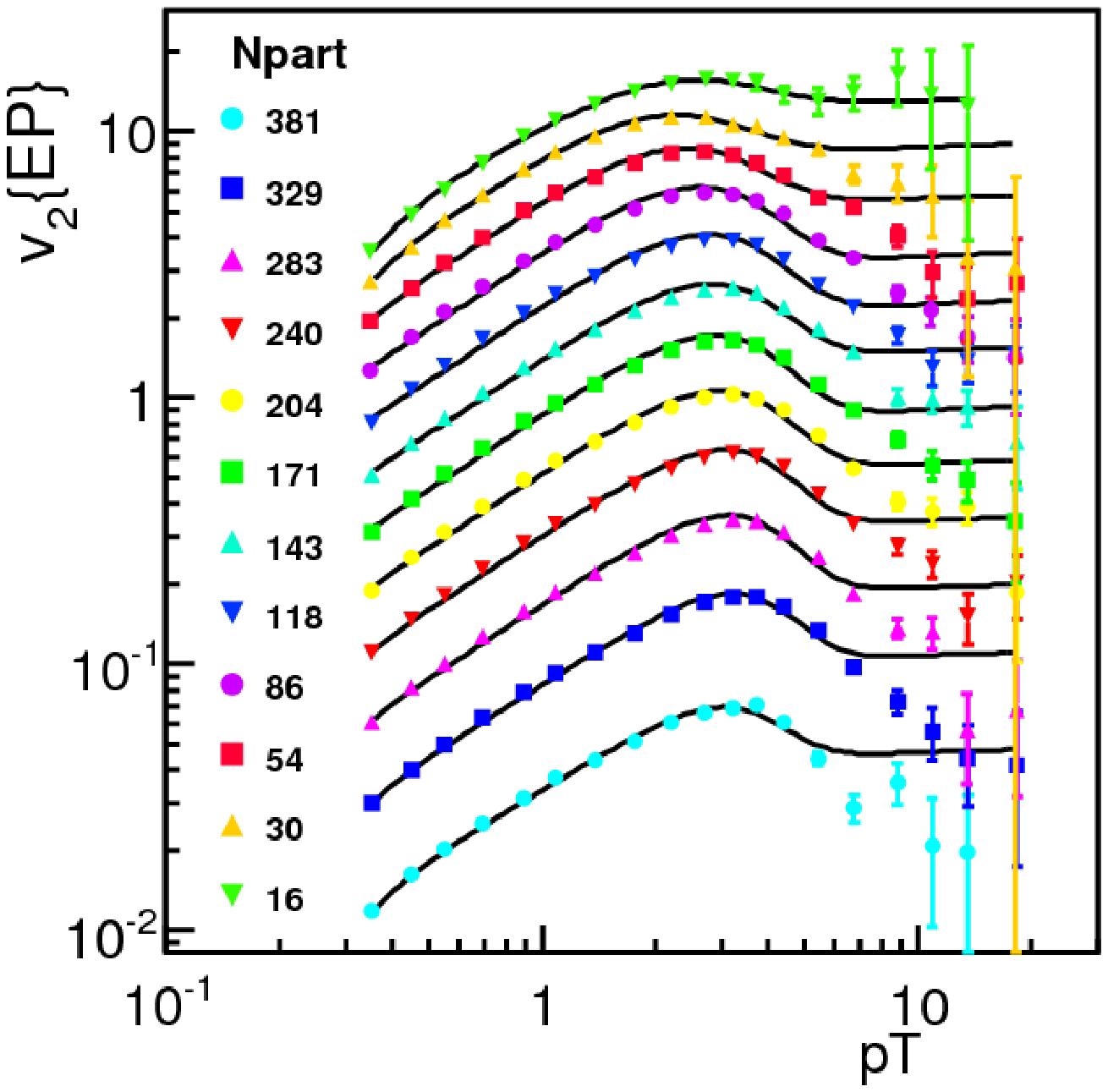} 
\includegraphics[width=0.45\textwidth]{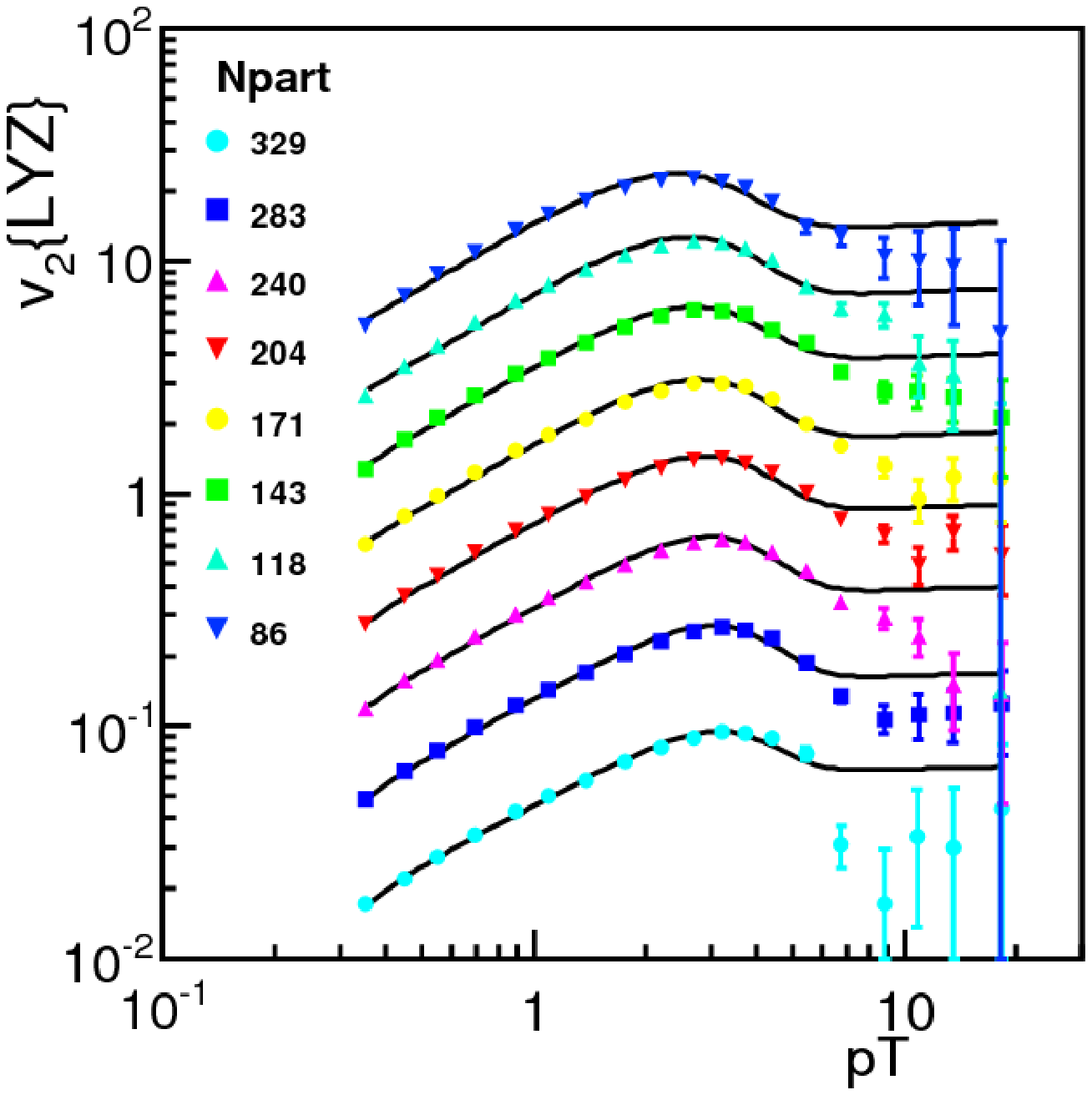} 
\end{center}
\caption{Fits of Eqs.~(\ref{eq17})~--~(\ref{eq18}) to CMS data on $v_2$ \cite{bib:cmsv2} in the case of four types of methods (event plane $v_2\lbrace EP\rbrace$, $2^{nd}$ and $4^{th}$ order cumulant $v_2\lbrace2\rbrace$ and $v_2\lbrace4\rbrace$ and Lee-Yang zeros $v_2\lbrace LYZ\rbrace$ methods). Fit parameters are plotted in Fig.~\ref{fig:v2param} and are enlisted in Tab.~\ref{tab:CMSv2param} \label{fig:v2}}
\end{figure}

\begin{figure}
\begin{center}
\includegraphics[width=0.45\textwidth, height=0.395\textheight]{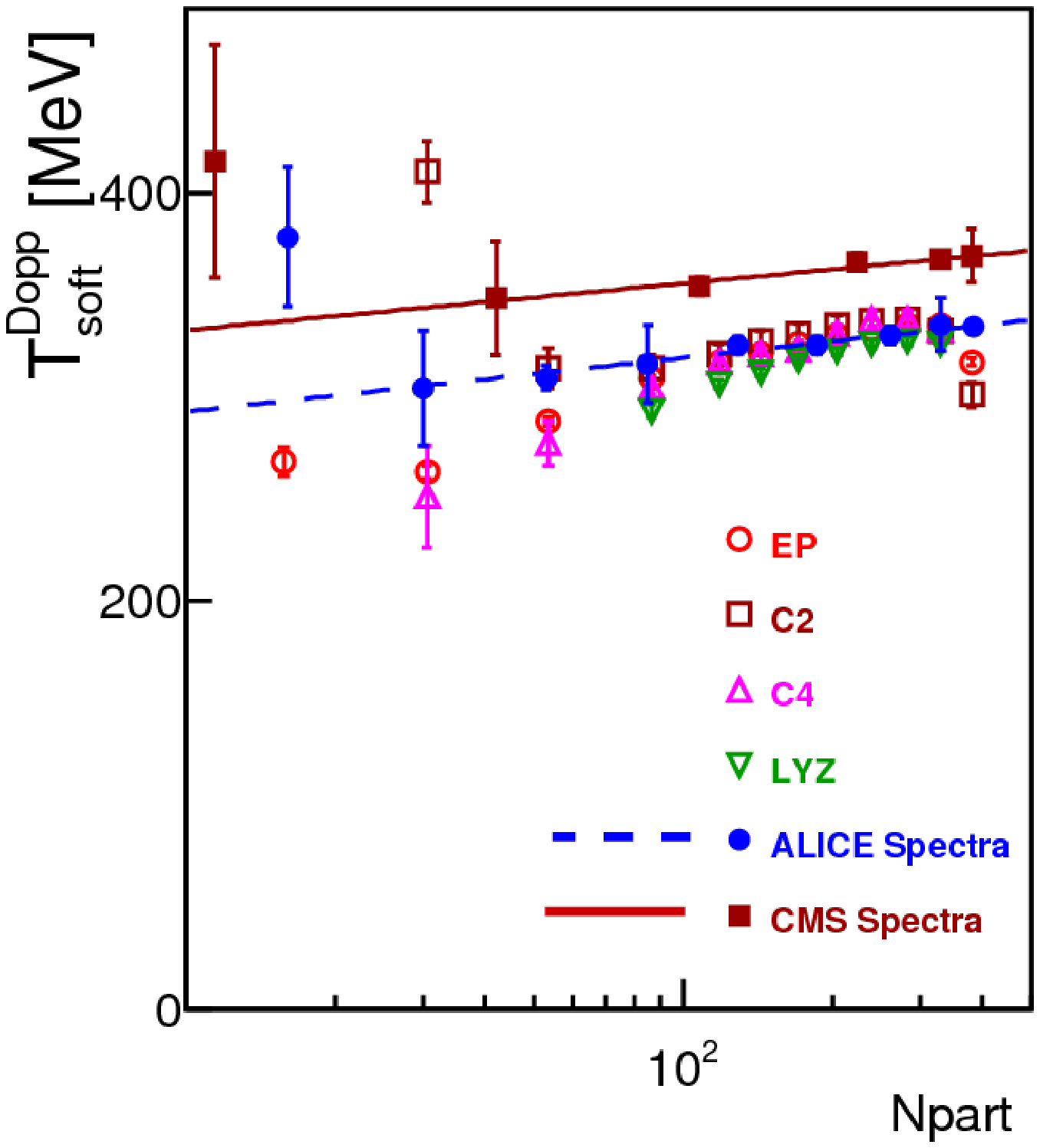}
\includegraphics[width=0.45\textwidth]{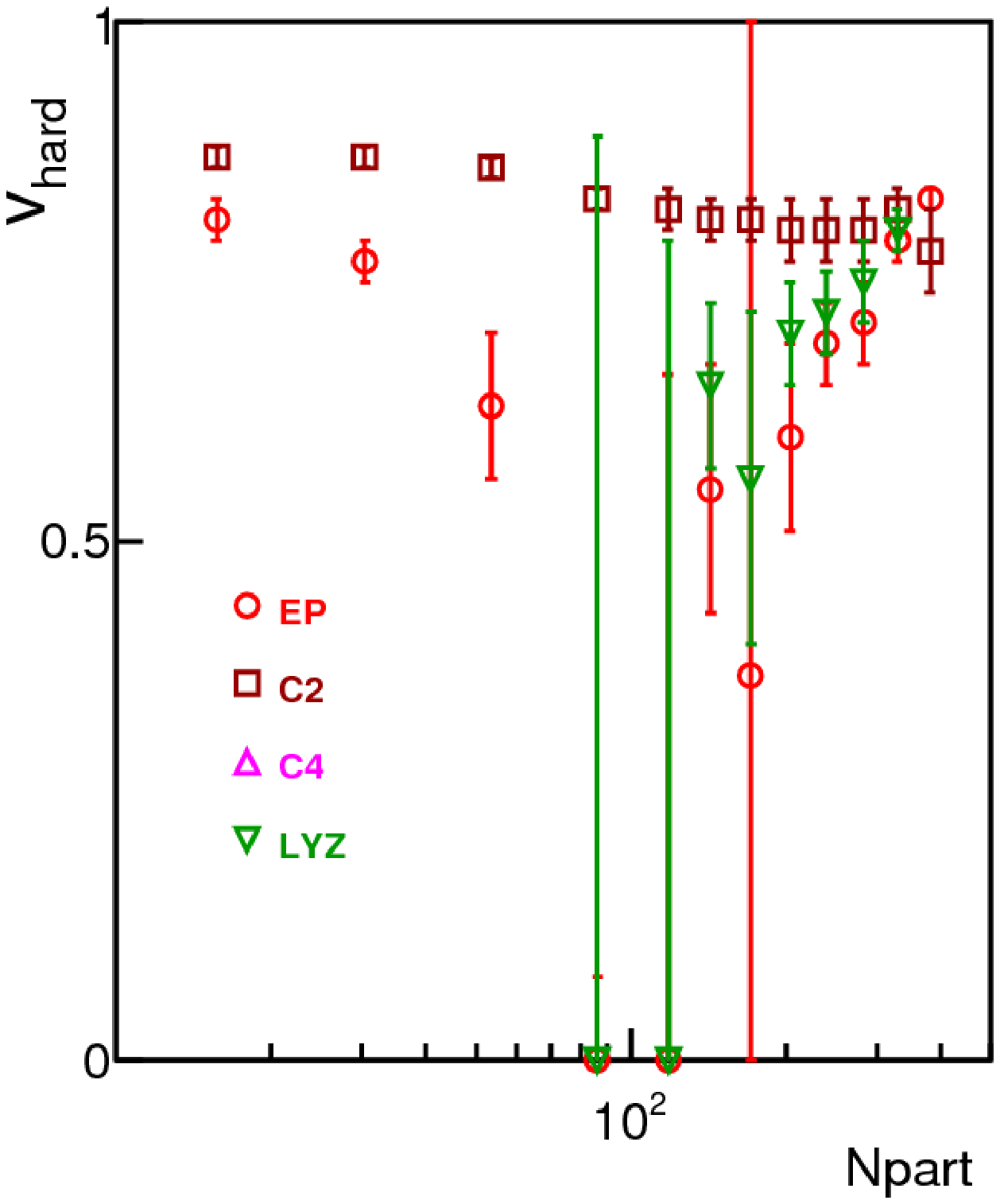}
\includegraphics[width=0.45\textwidth]{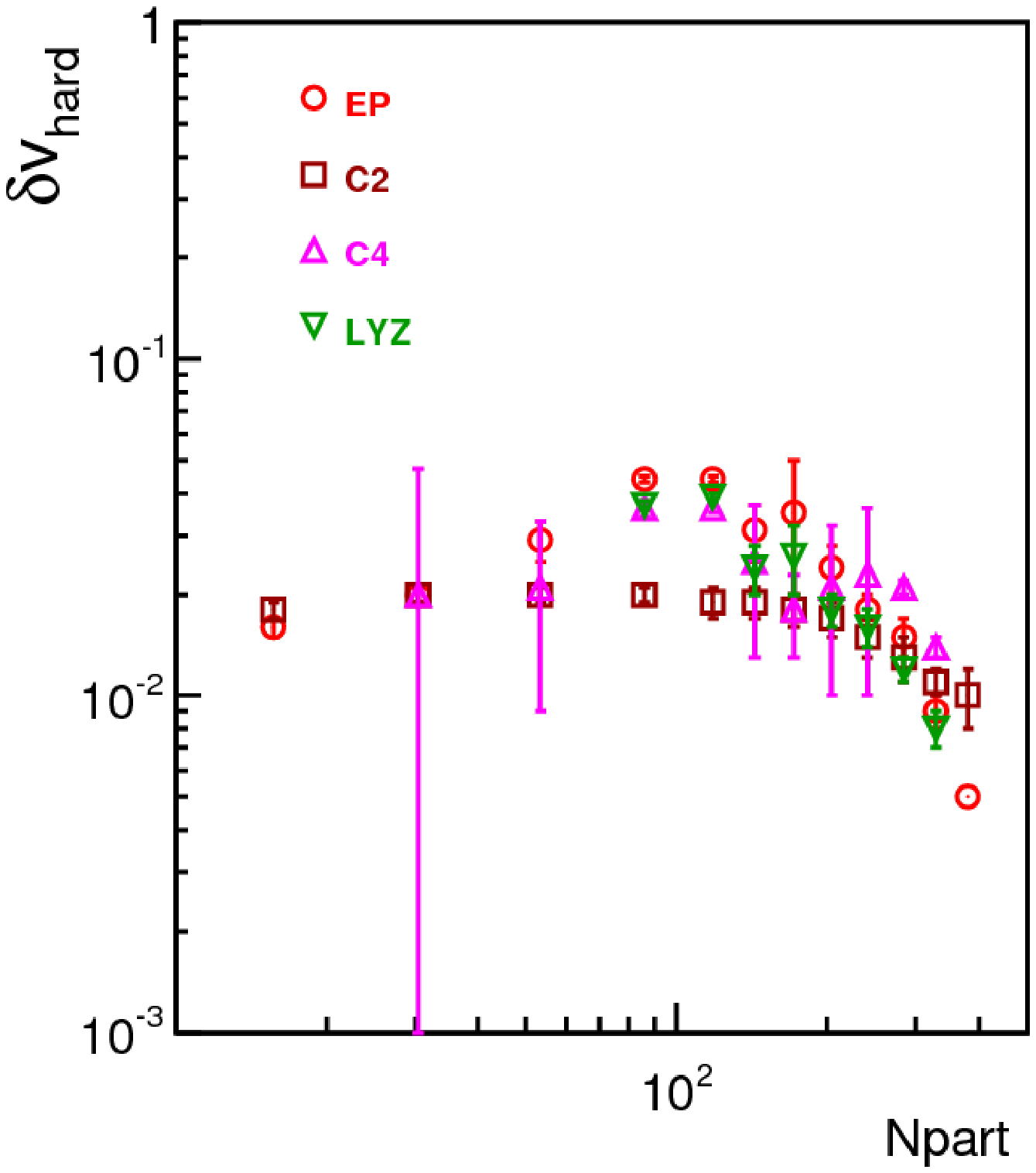}
\includegraphics[width=0.45\textwidth, height=0.395\textheight]{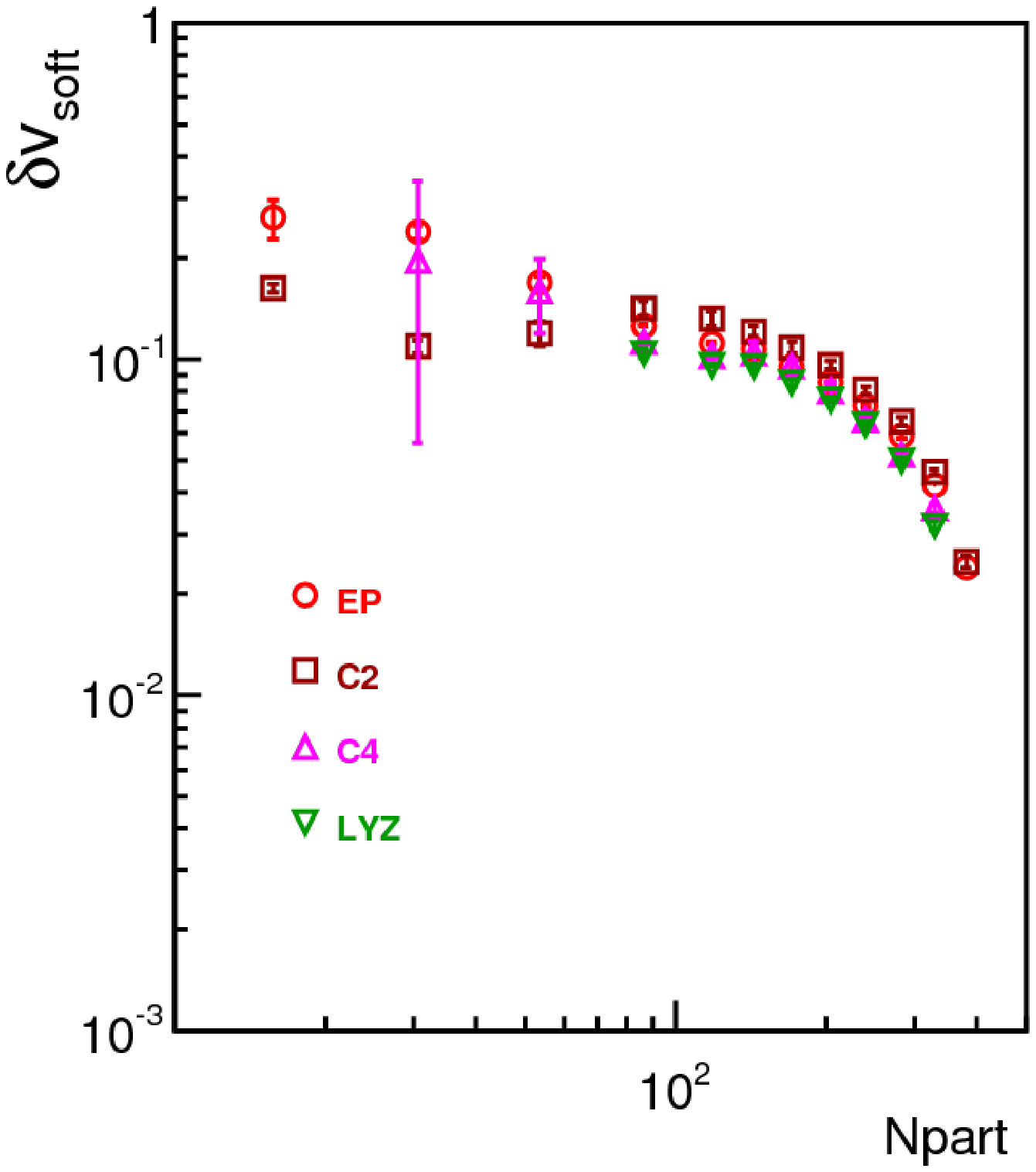}
\end{center}
\caption{Fitted parameters of Eqs.~(\ref{eq17})~--~(\ref{eq18}) to CMS data on $v_2$ \cite{bib:cmsv2} in the case of four types of methods (event plane ``$EP$'', $2^{nd}$ and $4^{th}$ order cumulant ``$C2$'' and ``$C4$'' and Lee-Yang zeros ``$LYZ$'' methods). Fits are shown in Fig.~\ref{fig:v2}. These fit parameters are enlisted in Tab.~\ref{tab:CMSv2param}\label{fig:v2param}. Bottom panels suggest that in this model, the anisotropy ($\delta v_{hard/soft}$) decreases for more central collisions.}
\end{figure}

Fits of Eqs.~(\ref{eq17})~--~(\ref{eq18}) to CMS data \cite{bib:cmsv2} on $v_2$ are found in Fig.~\ref{fig:v2}. The four different methods used in \cite{bib:cmsv2} for the extraction of $v_2$ are the $2^{nd}$ and $4^{th}$ order cumulant methods denoted by $v_2\lbrace2\rbrace$ and $v_2\lbrace4\rbrace$, the event-plane $v_2\lbrace EP\rbrace$ and Lee--Yang zeros $v_2\lbrace LYZ\rbrace$ methods. Fitted parameters are listed in Table~\ref{tab:CMSv2param}, and shown in Fig.~\ref{fig:v2param}.

When fitting $v_2$ data, all the parameters determined from fits to transverse spectra in Sec.~\ref{sec:fitspec} have been taken fixed except for the $T^{Dop}_{soft}$-s. As can be seen in the top-left panel of Fig.~\ref{fig:v2param}, $T^{Dop}_{soft}$-s obtained from fits to CMS $v_2$ data \cite{bib:cmsv2} are consistent with those obtained from fits to ALICE spectra \cite{bib:ALICEdNdpT}. 

Interestingly, $v_2$ data were found to be sensitive to the isotropic part of the flow velocity of the hard yields $v_{hard}$ (top-right panel of Fig.~\ref{fig:v2param}). In the case of $v_2\lbrace2\rbrace$, this parameter takes a quite high value of $v_{hard}\sim$ 0.85 (due to huge errorbars, $v_{hard}$ values extracted from the other methods are not conclusive). In the language of statistical hadronisation models, $v_{hard}$ means the average velocity of hadrons stemming from jets.

Finally, all four methods for the extraction of $v_2$ in \cite{bib:cmsv2} suggest that $\delta v_{hard}$ and $\delta v_{soft}$ (the $2^{nd}$ Fourier components of the transverse flow of the hard and the soft yields) decrease for more central collisions. This observation is in accordance with smaller anisotropy in more central collisions.

\section{Summary}
\label{sec:conc}
In this paper, we have \textit{simultanously} reproduced the transverse spectra \textit{and} the azimuthal anisotropy ($v_2$) of charged hadrons stemming from various centrality $PbPb$ collisions at $\sqrt s$ = 2.76 ATeV. In the proposed model, the hadron spectrum is assumed to be simply the sum of yields originated from ``soft'' and ``hard'' processes, Eq.~(\ref{eq1}). It is conjectured that hadrons are distributed according to the Tsallis distribution in both types of yields. As for the hard yields, this assumption is supported by the observation that the Tsallis distribution provides a reasonably good approximation for pion spectra obtained via $pQCD$-improved parton model calculations for central or peripheral $PbPb$ collisions at LHC energy (Fig.~\ref{fig:pQCD}). Furthermore, after the subtraction of the hard part fitted to the spectra, the remaining soft part can be described by another Tsallis distribution.

Analytic formulas have been obtained for the spectra and for $v_2$ in the limit of small transverse flow velocity fluctuations as a function of the azimuth angle. The parameters of the soft and hard Tsallis distributions have been determined from fits to transverse spectra and $v_2$ data measured by the CMS \cite{bib:cmsdNdpT,bib:cmsv2} and ALICE \cite{bib:ALICEdNdpT} collaborations. The dependence of the fitted parameters on the event centrality  ($N_{part}$) have been found similar in the case of the CMS and ALICE data. 
Fits to CMS data on $v_2$ suggest that in this model, the anisotropy decreases for more central collisions.


\section*{Acknowledgement}
\label{sec:ack}
This work was supported by Hungarian OTKA grants K104260, NK106119, and
NIH TET 12 CN-1-2012-0016. Author GGB also thanks the J\'anos Bolyai
Research Scholarship of the Hungarian Academy of Sciences.


\onecolumn


\begin{table}
\begin{center}
\begin{tabular}{ *{7}l }
  $N_{part}$ & centrality[\%] & $q_{hard}$        & $T^{Dop}_{hard}$ [MeV] & $q_{soft}$         & $T^{Dop}_{soft}$ [MeV] & $\chi^2/ndf$ \\[0.5mm] \hhline{=======} \noalign{\smallskip}
  381        &  $\;\;$0 -- 5  & 1.162 $\pm$ 0.001 & 158 $\pm$ 1            & 1.015 $\pm$ 0.006  & 369 $\pm$ 13           & 1.26     \\[0.5mm] \hline \noalign{\smallskip}  
  329        &  $\;\;$5 -- 10 & 1.160 $\pm$ 0.001 & 153 $\pm$ 4            & 1.017 $\pm$ 0.007  & 367 $\pm$ 2            & 0.8      \\[0.5mm] \hline \noalign{\smallskip}
  224        & 10 -- 30       & 1.158 $\pm$ 0.001 & 153 $\pm$ 5            & 1.019 $\pm$ 0.007  & 366 $\pm$ 4            & 1.22     \\[0.5mm] \hline \noalign{\smallskip}
  108        & 30 -- 50       & 1.155 $\pm$ 0.001 & 147 $\pm$ 4            & 1.026 $\pm$ 0.009  & 354 $\pm$ 4            & 0.73     \\[0.5mm] \hline \noalign{\smallskip}
$\;\;$42     & 50 -- 70       & 1.150 $\pm$ 0.001 & 142 $\pm$ 4            & 1.033 $\pm$ 0.012  & 348 $\pm$ 28           & 0.51     \\[0.5mm] \hline \noalign{\smallskip}
$\;\;$11     & 70 -- 90       & 1.145 $\pm$ 0.001 & 141 $\pm$ 3            & 1.015 $\pm$ 0.024  & 415 $\pm$ 57           & 0.56     \\[0.5mm] \hline \noalign{\smallskip}
\end{tabular}
\caption{Fitted parameters of Eqs.~(\ref{eq12})~--~(\ref{eq14}) to CMS data \cite{bib:cmsdNdpT} on transverse spectra of charged hadrons stemming from various centrality $PbPb$ collisions at $\sqrt{s}$ = 2.76 ATeV. Fits are shown in Fig.~\ref{fig:dNdpT}. $ndf$ = 21.\label{tab:CMSparam}}
\end{center}
\end{table}


\begin{table}
\begin{center}
\begin{tabular}{ *{7}l }
  $N_{part}$ & centrality[\%] & $q_{hard}$             & $T^{Dop}_{hard}$ [MeV] & $q_{soft}$              & $T^{Dop}_{soft}$ [MeV] & $\chi^2/ndf$ \\[0.5mm] \hhline{=======} \noalign{\smallskip}
  383        & $\;\;$0 -- 5   & 1.164 $\pm$ 0.001      & 115 $\pm$ 3            & 1.028 $\pm$ 0.006       & 334 $\pm$ $\;\;$3      & 0.39     \\[0.5mm] \hline \noalign{\smallskip}
  330        & $\;\;$5 -- 10  & 1.162 $\pm$ 0.001      & 121 $\pm$ 1            & 1.030 $\pm$ 0.006       & 335 $\pm$ 13           & 0.38     \\[0.5mm] \hline \noalign{\smallskip}
  261        & 10 -- 20       & 1.161 $\pm$ 0.001      & 125 $\pm$ 3            & 1.033 $\pm$ 0.006       & 330 $\pm$ $\;\;$4      & 0.29     \\[0.5mm] \hline \noalign{\smallskip}
  186        & 20 -- 30       & 1.160 $\pm$ 0.001      & 130 $\pm$ 3            & 1.037 $\pm$ 0.007       & 325 $\pm$ $\;\;$4      & 0.39     \\[0.5mm] \hline \noalign{\smallskip}
  129        & 30 -- 40       & 1.156 $\pm$ 0.001      & 138 $\pm$ 3            & 1.037 $\pm$ 0.008       & 325 $\pm$ $\;\;$4      & 0.23     \\[0.5mm] \hline \noalign{\smallskip}
  $\;\;$85   & 40 -- 50       & 1.154 $\pm$ 0.001      & 141 $\pm$ 1            & 1.042 $\pm$ 0.009       & 316 $\pm$ 19           & 0.30     \\[0.5mm] \hline \noalign{\smallskip}
  $\;\;$53   & 50 -- 60       & 1.152 $\pm$ 0.001      & 143 $\pm$ 2            & 1.046 $\pm$ 0.010       & 309 $\pm$ $\;\;$6      & 0.25     \\[0.5mm] \hline \noalign{\smallskip}
  $\;\;$30   & 60 -- 70       & 1.147 $\pm$ 0.002      & 148 $\pm$ 1            & 1.049 $\pm$ 0.014       & 304 $\pm$ 28           & 0.18     \\[0.5mm] \hline \noalign{\smallskip}
  $\;\;$16   & 70 -- 80       & 1.135 $\pm$ 0.001      & 166 $\pm$ 2            & 1.000 $\pm$ 0.050       & 378 $\pm$ 34           & 0.50     \\[0.5mm] \hline \noalign{\smallskip}
\end{tabular}
\caption{Fitted parameters of Eqs.~(\ref{eq12})~--~(\ref{eq14}) to ALICE data \cite{bib:ALICEdNdpT} on transverse spectra of charged hadrons stemming from various centrality $PbPb$ collisions at $\sqrt{s}$ = 2.76 ATeV. Fits are shown in Fig.~\ref{fig:dNdpT}. $ndf$ = 59.\label{tab:ALICEparam}}
\end{center}
\end{table}

\begin{table}[h!]
\begin{center}
\begin{tabular}{ c|*{4}l }
                      & $q_{2,soft}$      & $q_{2,hard}$      & $\mu_{soft}$       & $\mu_{hard}$       \\ \hhline{=|====}  \noalign{\smallskip}
CMS                   & 1.058 $\pm$ 0.025 & 1.136 $\pm$ 0.001 & -0.008 $\pm$ 0.005 & 0.005 $\pm$ 0.0003 \\ \hline  \noalign{\smallskip}
ALICE                 & 1.074 $\pm$ 0.018 & 1.130 $\pm$ 0.003 & -0.009 $\pm$ 0.004 & 0.006 $\pm$ 0.0006 \\ \hline  \noalign{\bigskip}
                      & $T^{soft}_{1}$ [MeV] & $T^{hard}_{1}$ [MeV] & $\tau_{soft}$ [MeV]   & $\tau_{hard}$ [MeV]     \\ \hhline{=|====}  \noalign{\smallskip}
CMS                   & 310 $\pm$ 2          & 126 $\pm$ 5          & 10 $\;\;\pm$ 3.7          & $\;\;\;$5.3 $\pm$ 0.8   \\ \hline  \noalign{\smallskip}
ALICE                 & 266 $\pm$ 2          & 194 $\pm$ 2          & 11.5 $\pm$ 3          & -12.5 $\pm$ 0.5         \\ \hline  \noalign{\smallskip}
\end{tabular}
\caption{Parameters of Eq.~(\ref{eq15}) obtained from fits shown in the bottom panels of Fig.~\ref{fig:dNdpT}.\label{tab:qTfit}}
\end{center}
\end{table}


\begin{table}
\begin{center}
\begin{tabular}{ c|*{8}l }
                      & $N_{part}$ & centrality[\%] & $v_{hard}$      & $\delta v_{hard}$ & $T^{Dop}_{soft}$ [MeV]  & $\delta v_{soft}$  & $\chi^2/ndf$     \\ \hhline{=|=======}      
\multirow{12}{*}{EP}  & 381        & $\;\;$0 -- 5   & 0.83 $\pm$ 0.01 & 0.005 $\pm$ 0.000 & 317 $\pm$ 2              & 0.024 $\pm$ 0.000 & $\;\;$5.8        \\ \cline{2-8} 
                      & 329        & $\;\;$5 -- 10  & 0.79 $\pm$ 0.02 & 0.009 $\pm$ 0.001 & 335 $\pm$ 2              & 0.042 $\pm$ 0.000 & $\;\;$8.4        \\ \cline{2-8} 
                      & 283        & 10 -- 15       & 0.71 $\pm$ 0.04 & 0.015 $\pm$ 0.002 & 335 $\pm$ 1              & 0.059 $\pm$ 0.001 & 12.9             \\ \cline{2-8} 
                      & 240        & 15 -- 20       & 0.69 $\pm$ 0.04 & 0.018 $\pm$ 0.002 & 333 $\pm$ 1              & 0.073 $\pm$ 0.001 & 10.9             \\ \cline{2-8} 
                      & 204        & 20 -- 25       & 0.60 $\pm$ 0.09 & 0.024 $\pm$ 0.004 & 330 $\pm$ 1              & 0.085 $\pm$ 0.001 & 13.2             \\ \cline{2-8} 
                      & 171        & 25 -- 30       & 0.37 $\pm$ 0.67 & 0.035 $\pm$ 0.015 & 326 $\pm$ 1              & 0.095 $\pm$ 0.005 & 13.0             \\ \cline{2-8} 
                      & 143        & 30 -- 35       & 0.55 $\pm$ 0.12 & 0.031 $\pm$ 0.006 & 321 $\pm$ 1              & 0.107 $\pm$ 0.003 & 12.3             \\ \cline{2-8} 
                      & 118        & 35 -- 40       & 0.00 $\pm$ 0.66 & 0.044 $\pm$ 0.001 & 317 $\pm$ 1              & 0.111 $\pm$ 0.001 & 10.9             \\ \cline{2-8} 
                      & 86.2       & 40 -- 50       & 0.00 $\pm$ 0.08 & 0.044 $\pm$ 0.001 & 309 $\pm$ 1              & 0.126 $\pm$ 0.001 & 13.9             \\ \cline{2-8} 
                      & 53.5       & 50 -- 60       & 0.63 $\pm$ 0.07 & 0.029 $\pm$ 0.004 & 288 $\pm$ 2              & 0.169 $\pm$ 0.007 & $\;\;$8.1        \\ \cline{2-8} 
                      & 30.5       & 60 -- 70       & 0.77 $\pm$ 0.02 & 0.020 $\pm$ 0.001 & 263 $\pm$ 3              & 0.238 $\pm$ 0.011 & $\;\;$2.8        \\ \cline{2-8} 
                      & 15.7       & 70 -- 80       & 0.81 $\pm$ 0.02 & 0.016 $\pm$ 0.001 & 268 $\pm$ 7              & 0.263 $\pm$ 0.035 & $\;\;$0.4        \\ \hhline{=|=======}      

\multirow{12}{*}{C2}  & 381        & $\;\;$0 -- 5   & 0.78 $\pm$ 0.04 & 0.010 $\pm$ 0.002 & 301 $\pm$ 6              & 0.025 $\pm$ 0.001 & $\;\;$2.0        \\ \cline{2-8} 
                      & 329        & $\;\;$5 -- 10  & 0.82 $\pm$ 0.02 & 0.011 $\pm$ 0.001 & 332 $\pm$ 4              & 0.046 $\pm$ 0.001 & $\;\;$1.5        \\ \cline{2-8} 
                      & 283        & 10 -- 15       & 0.80 $\pm$ 0.03 & 0.013 $\pm$ 0.002 & 338 $\pm$ 5              & 0.065 $\pm$ 0.002 & $\;\;$0.8        \\ \cline{2-8} 
                      & 240        & 15 -- 20       & 0.80 $\pm$ 0.03 & 0.015 $\pm$ 0.002 & 337 $\pm$ 5              & 0.081 $\pm$ 0.002 & $\;\;$1.5        \\ \cline{2-8} 
                      & 204        & 20 -- 25       & 0.80 $\pm$ 0.03 & 0.017 $\pm$ 0.002 & 335 $\pm$ 5              & 0.096 $\pm$ 0.003 & $\;\;$1.1        \\ \cline{2-8} 
                      & 171        & 25 -- 30       & 0.81 $\pm$ 0.02 & 0.018 $\pm$ 0.002 & 331 $\pm$ 5              & 0.108 $\pm$ 0.004 & $\;\;$1.3        \\ \cline{2-8} 
                      & 143        & 30 -- 35       & 0.81 $\pm$ 0.02 & 0.019 $\pm$ 0.002 & 327 $\pm$ 5              & 0.121 $\pm$ 0.004 & $\;\;$0.8        \\ \cline{2-8} 
                      & 118        & 35 -- 40       & 0.82 $\pm$ 0.02 & 0.019 $\pm$ 0.002 & 321 $\pm$ 5              & 0.132 $\pm$ 0.006 & $\;\;$0.7        \\ \cline{2-8} 
                      & 86.2       & 40 -- 50       & 0.83 $\pm$ 0.01 & 0.020 $\pm$ 0.001 & 314 $\pm$ 6              & 0.142 $\pm$ 0.007 & $\;\;$1.8        \\ \cline{2-8} 
                      & 53.5       & 50 -- 60       & 0.86 $\pm$ 0.01 & 0.020 $\pm$ 0.001 & 314 $\pm$ 10             & 0.119 $\pm$ 0.010 & $\;\;$2.8        \\ \cline{2-8} 
                      & 30.5       & 60 -- 70       & 0.87 $\pm$ 0.01 & 0.020 $\pm$ 0.001 & 410 $\pm$ 15             & 0.109 $\pm$ 0.005 & $\;\;$5.8        \\ \cline{2-8} 
                      & 15.7       & 70 -- 80       & 0.87 $\pm$ 0.01 & 0.018 $\pm$ 0.001 & 500 $\pm$ 490            & 0.162 $\pm$ 0.005 & $\;\;$1.9        \\ \hhline{=|=======}      

\multirow{10}{*}{C4}  & 329        & $\;\;$5 -- 10  & 0.00 $\pm$ 0.96 & 0.014 $\pm$ 0.001 & 333 $\pm$ 5              & 0.036 $\pm$ 0.001 & $\;\;$2.4        \\ \cline{2-8} 
                      & 283        & 10 -- 15       & 0.00 $\pm$ 0.98 & 0.021 $\pm$ 0.001 & 338 $\pm$ 4              & 0.052 $\pm$ 0.001 & $\;\;$3.2        \\ \cline{2-8} 
                      & 240        & 15 -- 20       & 0.39 $\pm$ 0.69 & 0.023 $\pm$ 0.013 & 338 $\pm$ 4              & 0.066 $\pm$ 0.004 & $\;\;$4.2        \\ \cline{2-8} 
                      & 204        & 20 -- 25       & 0.56 $\pm$ 0.30 & 0.021 $\pm$ 0.011 & 331 $\pm$ 4              & 0.080 $\pm$ 0.005 & $\;\;$2.2        \\ \cline{2-8} 
                      & 171        & 25 -- 30       & 0.71 $\pm$ 0.10 & 0.018 $\pm$ 0.005 & 323 $\pm$ 4              & 0.095 $\pm$ 0.005 & $\;\;$1.5        \\ \cline{2-8} 
                      & 143        & 30 -- 35       & 0.57 $\pm$ 0.29 & 0.025 $\pm$ 0.012 & 321 $\pm$ 4              & 0.104 $\pm$ 0.009 & $\;\;$2.2        \\ \cline{2-8} 
                      & 118        & 35 -- 40       & 0.00 $\pm$ 0.98 & 0.036 $\pm$ 0.001 & 317 $\pm$ 4              & 0.102 $\pm$ 0.002 & $\;\;$3.5        \\ \cline{2-8} 
                      & 86.2       & 40 -- 50       & 0.00 $\pm$ 0.98 & 0.036 $\pm$ 0.002 & 305 $\pm$ 4              & 0.113 $\pm$ 0.004 & $\;\;$1.6        \\ \cline{2-8} 
                      & 53.5       & 50 -- 60       & 0.64 $\pm$ 0.27 & 0.021 $\pm$ 0.012 & 277 $\pm$ 11             & 0.159 $\pm$ 0.039 & $\;\;$1.2        \\ \cline{2-8} 
                      & 30.5       & 60 -- 70       & 0.57 $\pm$ 0.78 & 0.020 $\pm$ 0.027 & 251 $\pm$ 25             & 0.197 $\pm$ 0.141 & $\;\;$1.2        \\ \hhline{=|=======}      

\multirow{8}{*}{LYZ}  & 329        & $\;\;$5 -- 10  & 0.80 $\pm$ 0.02 & 0.008 $\pm$ 0.001 & 326 $\pm$ 3              & 0.032 $\pm$ 0.001 & $\;\;$4.7        \\ \cline{2-8} 
                      & 283        & 10 -- 15       & 0.75 $\pm$ 0.04 & 0.012 $\pm$ 0.001 & 327 $\pm$ 2              & 0.050 $\pm$ 0.001 & $\;\;$5.0        \\ \cline{2-8} 
                      & 240        & 15 -- 20       & 0.72 $\pm$ 0.04 & 0.016 $\pm$ 0.002 & 326 $\pm$ 2              & 0.064 $\pm$ 0.001 & $\;\;$5.0        \\ \cline{2-8} 
                      & 204        & 20 -- 25       & 0.70 $\pm$ 0.05 & 0.018 $\pm$ 0.002 & 322 $\pm$ 2              & 0.076 $\pm$ 0.001 & $\;\;$5.8        \\ \cline{2-8} 
                      & 171        & 25 -- 30       & 0.56 $\pm$ 0.16 & 0.026 $\pm$ 0.006 & 318 $\pm$ 2              & 0.085 $\pm$ 0.002 & $\;\;$7.1        \\ \cline{2-8} 
                      & 143        & 30 -- 35       & 0.65 $\pm$ 0.08 & 0.024 $\pm$ 0.004 & 312 $\pm$ 2              & 0.095 $\pm$ 0.002 & $\;\;$5.8        \\ \cline{2-8} 
                      & 118        & 35 -- 40       & 0.00 $\pm$ 0.79 & 0.039 $\pm$ 0.001 & 306 $\pm$ 2              & 0.096 $\pm$ 0.001 & $\;\;$7.4        \\ \cline{2-8} 
                      & 86.2       & 40 -- 50       & 0.00 $\pm$ 0.89 & 0.037 $\pm$ 0.001 & 293 $\pm$ 3              & 0.104 $\pm$ 0.003 & $\;\;$2.0        \\ \hline      

\end{tabular}
\caption{Fitted parameters of Eqs.~(\ref{eq17})~--~(\ref{eq18}) to CMS data on $v_2$ \cite{bib:cmsv2} in the case of four types of methods (event plane ``$EP$'', $2^{nd}$ and $4^{th}$ order cumulant ``$C2$'' and ``$C4$'', and Lee-Yang zeros ``$LYZ$'' methods). Fits are shown in Fig.~\ref{fig:v2}. These fit parameters are plotted in Fig.~\ref{fig:v2param}. $ndf$ = 15.\label{tab:CMSv2param}}
\end{center}
\end{table}

\twocolumn{}





\end{document}